\documentclass[
    ,final            
  ]
  {aipproc}

\layoutstyle{6x9}

\begin{document}

\title{Polarization in Heavy Ion Physics}


\author{S.M. Troshin}{address=
{Institute for High Energy Physics, Protvino, Moscow Region,
142281, Russia}}

\begin{abstract}
Role of polarization studies in heavy ion physics is discussed
with emphasis on the search for quark-gluon plasma formation and
studies of its dynamical properties.
\end{abstract}

\maketitle


\section{Introduction}
The main goal of the experiments performed with heavy ion collisions
is generation of the deconfined state of the QCD matter and studies of its properties.
Existence of deconfined
state was predicted prior to the QCD era (cf. e.g. \cite{mcler} and references
therein).
Asymptotic freedom of QCD has  led straightforwardly to the prediction that the
deconfined state should  be a weakly bounded state of quarks and gluons.
Important tools in the studies of the nature of the new form of
matter are the anisotropic flows which are the quantitative
characteristics of the collective motion of the produced hadrons
in the nuclear interactions. With their measurements one can
obtain a valuable information on  the early stages of reactions
and observe signals of QGP formation. Despite that the polarization measurements
in this field are complementary, they nevertheless
are  very useful tool for the searches for the deconfined state and
 study of its nature. Of course, in heavy--ion collisions spin states
 of final particles can only be measured.
\section{Polarization and search for quark-gluon plasma}
Assuming  different dynamics of hadron and nuclear
collisions, i.e. that high densities of matter reached in nuclear collisions
would lead to QGP formation,
 the idea to use vanishing polarization of $\Lambda$ hyperons
produced in nuclear collisions as a signal of QGP
has been exploited  in the early studies \cite{hoy,pan,stock}.

Indeed, in hadron interaction the experimental
 situation with hyperon polarization is widely known and stable for a long time.
Polarization of $\Lambda$ produced in the unpolarized inclusive $pp$--interactions
is negative and energy
independent. It increases linearly with $x_F$ at large transverse momenta
($p_\perp\geq 1$ GeV/c),
and for such
values of transverse momenta   is almost
$p_\perp$-independent, this dependence is represented in Fig. 1.
\begin{figure}[hbt]
\resizebox{8cm}{!}{\includegraphics*{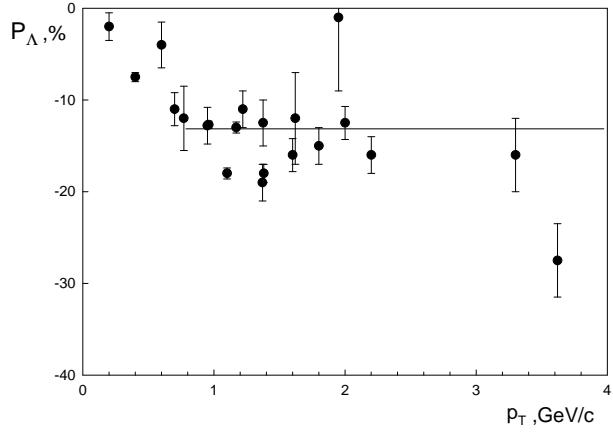}}
\caption{Transverse momentum dependence of polarization of $\Lambda$.}
\end{figure}
It should be noted that polarization was measured with respect to the production
plane, i.e. the plane spanned over the vector of the initial momentum
and the vector of the produced particle momentum. It was supposed that in the nuclear
collisions the produced transient state has an isotropic distribution of parton momenta
and therefore QGP lose any memory of the initial state momenta. The only existing vector
is the vector of the final state momentum. But due to parity conservation polarization along
this direction should be zero. Thus, the absence of hyperon polarization w.r.t. production
plane  was predicted to be a signal of QGP formation.
Vanishing hyperon polarization w.r.t. production plane in central nuclear collisions
  follows  from various models also.

However,  there are no experimental measurements at high energies
   of hyperon
polarization w.r.t. \textbf{production} plane in nuclear collisions and this prediction is still
a hypothesis.

\begin{figure}[hbt]
\resizebox{6cm}{!}{\includegraphics*{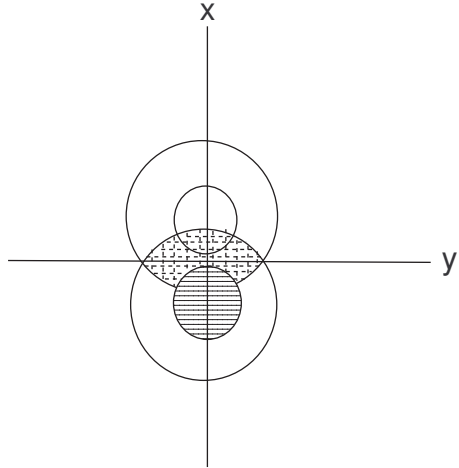}}
\caption{Reaction plane in hadron and nuclei collisions.}
\end{figure}
Here we turn to another aspect of the same problem. For this purpose it is important to
make a remark
 that in the recent  paper \cite{weiner}, which
 provides an emphasis on the historic aspects of the QGP
searches, the  conclusion made that the  deconfined state of matter has
being observed in hadronic reactions during a long time and it would be
interesting to study collective properties of transient state in
reactions with hadrons and nuclei simultaneously since the dynamics is qualitatively
the same and differences are merely quantitative.
Another array of theoretical predictions is related to prediction of significant
polarization
w.r.t. \textbf{reaction} plane in the non-central hadron and nuclear
collisions dew to the presence of the large orbital angular
momentum in such collisions. Reaction plane is the plane spanned over the vectors
of the initial momentum $\vec{p}$ and the  impact parameter of colliding particles
or nuclei $\vec{b}$. This definition is illustrated in  Fig. 2, where the normal
 to the reaction plane is directed along the $y$ axis.

The
determination of the reaction plane in the non-central hadronic collisions could be
 experimentally feasible  with use of the standard procedure.
The relationship of the impact parameter with the
 final state multiplicity is a useful tool in these studies similar to the studies
 of the nuclei interactions.
The value of impact parameter can be  determined through
 the centrality and then, e.g. global polarization, directed  or elliptic flow
  can be analyzed by selecting events
 in a specific centrality ranges.

It is evident that orbital angular momentum can have very large values. It can be estimated
as
 follows
\begin{equation}\label{l}
 L(s,b) \simeq \alpha b \frac{\sqrt{s}}{2}D_C(b),
\end{equation}
where $D_C(b)$ describes distribution of matter in the overlap region (cf. Fig.~2).
It should be noted that $L\to 0$ at $b\to\infty$ and $L=0$ at $b=0$ (Fig.~3).
\begin{figure}[hbt]
\resizebox{6cm}{!}{\includegraphics*{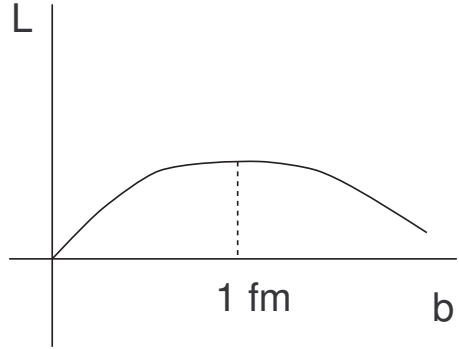}}
\caption{Qualitative dependence of the orbital angular momentum $L$ on the impact parameter $b$.}
\end{figure}
In nuclear collisions at RHIC and LHC average values of orbital angular momentum are of order
($\hbar=1$) $10^5$ (RHIC) and $10^7$ (LHC)\footnote{In nuclear collisions orbital angular momentum
has a maximum at $b\simeq R_A$.}.

It is natural that the question arises:
what are  the observable effects of the large
angular orbital momentum presented in the peripheral heavy-ion and hadron collisions?
Carruthers supposed that orbital angular momentum would be released in the
coherent burst of polarized vector mesons \cite{carru83}.
Yang and Chou pointed out to the  strong necessity for the spins of the outgoing
particles to line up parallel to each other in the
transverse direction to the reaction plane \cite{chyan86},
i.e. $(\sigma_1^T\cdot\sigma_2^T)_{average}>0$ due to a net deficiency
in the left and right-moving outgoing systems.
The reaction plane is  perpendicular to the orbital angular momentum direction.
Thus, in a process of high energy hadron and nuclear collisions large initial
orbital angular momentum can, in principle, be converted into the spin angular momentum of
final particles resulting in their  polarization relative to the reaction plane.
We will try to connect this possibility with the nature of transient strong interaction matter.

\section{Polarization and dynamics of quark-gluon plasma}
Weakly-coupled matter of QGP (parton model with final state interactions)
does not allow coherent collective rotation of the system and therefore
 finite transverse gradient of the
 average longitudinal momentum per produced parton should exist in the overlap region.
 It is claimed that relative OAM in collision of partons will lead to global
  quark polarization due to spin-orbital coupling \cite{liang}.

Significant (order of $0.3$) polarization
of hyperons relative to reaction plane was anticipated due to global polarization
of quarks.
Similar ideas were used for the hyperon polarization in hadron non-central collisions \cite{volosh}.
The idea to observe circularly polarized direct photons as a signal
  of quark polarization in the QGP was proposed in \cite{ipp,betz}.

  The measurements of global polarization  $\Lambda$ and $\bar\Lambda$
  were performed at RHIC (STAR Collaboration) in $Au+Au$ collisions
  at $\sqrt{s_{NN}}= 62.4$ and $200$ GeV and upper limit $|P_{\Lambda,\bar\Lambda}|\leq 0.02$
has been obtained \cite{abelev}. It should be noted that
global spin alignment for $\phi$ and $K^{*0}$ were also not observed for the
 different centralities \cite{selyu}. Thus, if we will not consider scenario of
  complete dilution of polarization
  by  hadronization mechanism, we should conclude that at the moment
  no experimental evidence
 exists for conversion of the orbital angular momentum into the spin angular
  momentum in nuclear collisions and this conclusion is correlated but not
   necessarily follow from the result on the strongly interacting nature
   of transient matter observed at RHIC.

It is well known that discovery of the deconfined state of matter has been announced  by the four
 major experiments at RHIC
Despite the highest values of energy and density have been reached,
a genuine quark-gluon plasma QGP was not found. The deconfined state reveals
 the properties of the perfect liquid, being strongly interacting collective
  state and therefore it was labelled as sQGP.
  The question  arise again: what are the experimental manifestations of the
   large orbital angular momentum could be in the case of strongly interacting transient matter?

\begin{figure}[hbt]
\resizebox{6cm}{!}{\includegraphics*{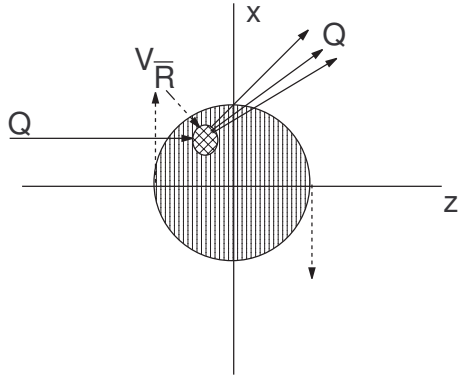}}
\caption{Interaction of the constituent quark with rotating quark-pion liquid.}
\end{figure}

 First, observation of an ideal liquid as a transient state  means that
  this state of matter has low viscosity, i.e.
  large interaction cross-section (estimates provide the value about 22 mb).
  Then we should conclude that in these circumstances the preexisting
   large orbital angular momentum  will lead to rotation of the transient deconfined
   matter in the overlap region as a whole, i.e. all parts have the same angular velocity
    and the orbital angular momentum does not convert to the spin angular momentum.
This rotation will lead to the directed flow $v_1$ as it was shown in \cite{ttint}.
Assumed there particle production mechanism at moderate transverse
momenta is an excitation of  a part of the rotating transient state
 of  massive constituent quarks (interacting by pion exchanges) by
 the one of the valence constituent quarks with  subsequent hadronization of the quark-pion liquid droplets
 (cf. Fig. 4).
  The directed flow $v_1$ in nuclei collisions as well as
in hadron reactions  depends
on  the rapidity difference $y-y_{beam}$ and not on the incident energy.
The mechanism
 therefore can provide a qualitative explanation of the incident-energy scaling \cite{v1exp,v11,v12,v13}
 of $v_1$ observed at RHIC (Fig.5).
\begin{figure}[hbt]
\resizebox{8cm}{!}{\includegraphics*{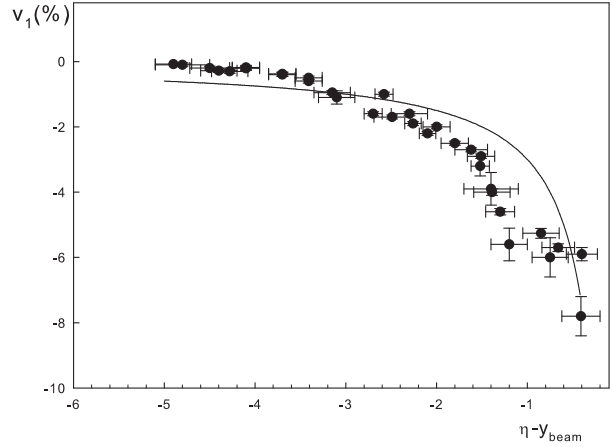}}
\caption{Dependence of directed flow on $\eta-y_{beam}$ in $Au+Au$ and $Cu+Cu$ collisions
at 62.4 GeV and 200 GeV at RHIC. }
\end{figure}

 Thus, the strongly interacting nature of transient state
would lead to the non-zero directed flow $v_1$ and vanishing global polarization.
\section{Expectations for the LHC energies}
The evident question nowadays is about the nature of the matter which will be produced at the
LHC.

If the vanishing directed flow $v_1$ would be observed then one could conclude on
the weakly-coupled system, or genuine QGP. But then what about global polarization?
Is this a signal of QGP formation? Not necessarily, and the reason is the
existence of the reflective (antishadow) scattering at the LHC energies.
This particular mechanism is not
related to the question on what kind of deconfined matter was produced \cite{imb,imb1}.
The  geometric picture at the energy which is beyond the
black disc limit can be described as a scattering off the partially
 reflective and the partially absorptive disk surrounded by the black ring which turns out
 into a grey area at larger values of the impact parameter.
The evolution with energy  is characterized by increasing albedo due to the
 interrelated  increase of reflection  and decrease of absorption at small impact parameters.
 This mechanism results from unitarity saturation and leads to the peripheral production of
 secondary particles in the impact parameter space. The emerging global polarization of produced
 particles will appear then due to the imbalance of the orbital angular momentum in the initial and
 final states. Indeed, the enhancement of the peripheral particle production
  would destroy the balance
 between orbital angular momentum in the initial and final states; most of the particles
 in the final state would carry  out  significant orbital
 angular momentum.
To compensate this orbital momentum the spins of the secondary particles should  become
  lined up.
\section{Conclusion}
Concluding on the polarization studies in heavy ion physics, it is evident that
 there are more questions than answers   in this field at the moment.
Nevertheless, there are no doubts  that polarization measurements combined
with anisotropic flow measurements will be able to help in detecting
 QGP and revealing   its dynamical properties more unambiguously.
At  RHIC there is another possibility of the directed
flow measurements in the polarized proton collisions.
As it was discussed above, the magnitudes of global polarization and
 directed flow are in a prompt relation with the nature of the transient matter,
namely, depending on whether it is weakly or strongly coupled matter, these observables
have different values. The  role of polarization measurements in heavy ion
physics certainly deserves further studies --- both experimental and theoretical.
\begin{theacknowledgments}
  Fruitful discussions with N.~E. Tyurin and support are gratefully acknowledged. I am also grateful
  to D.~G. Crabb for invitation and  warm hospitality at the University of Virginia  during Symposium  SPIN 2008.
\end{theacknowledgments}

\end{document}